\documentclass[review]{elsarticle}
\usepackage{lineno,hyperref}
\modulolinenumbers[5]
\journal{Journal of \LaTeX\ Templates}

\begin{document}
\begin{frontmatter}

\title{Topological Mott transition in a two band model of spinless fermions  with on-site Coulomb repulsion}
\tnotetext[mytitlenote]{Topological Mott transition in a two band model of spinless fermions  with on-site Coulomb repulsion}
\author{Igor N.Karnaukhov}
\address{G.V. Kurdyumov Institute for Metal Physics, 36 Vernadsky Boulevard, 03142 Kiev, Ukraine}
\fntext[myfootnote]{karnaui@yahoo.com}



\begin{abstract}
In the framework of mean field approach, we study topological Mott transition in a two band model of spinless fermions on a  square lattice at half filling. We consider the combined effect of the on-site Coulomb repulsion and the spin-orbit  Rashba coupling.
The ground state phase diagram is calculated as a function of the strength of the spin-orbit Rashba coupling and the Coulomb repulsion.
The spin-orbit Rashba coupling leads to a distinct phase of matter, the topological semimetal.
We study a new type of phase transition between the non-topological insulator  and  topological semimetal states. Topological phase state is characterized by the zero energy Majorana states, which there are in defined region of the wave vectors and are localized at the boundaries of the sample. The region of existence of the zero energy Majorana states tends to zero at the point of the Mott phase transition. The zero energy Majorana states are dispersionless (they can be considered as flat bands), the Chern number and Hall conductance are equal to zero (note in two dimensional model)
\end{abstract}

\begin{keyword}
\texttt Hubbard model  \sep zero energy Majorana states \sep Mott transition
\end{keyword}
\end{frontmatter}

\section{Introduction}

The experimental discovery of the first Weyl semimetal in TaAs \cite{A1} gave impetus to the study of topological gapless states in real compounds.
The chiral surface states  determine the topological features of the phase state: in one-dimensional systems, these are the zero energy Majorana states \cite{Kit1,a3,K3} (the Majorana fermions), in two-dimensional systems, we are talking about  chiral edge modes localized at the boundaries \cite{Kit2,2D} (the Dirac fermions), in three-dimensional systems there are the Weyl nodes and topological surface states \cite{A1,3D1,3D2} (the Weyl fermions). As a rule, topological surface states of fermions exist for a certain dimension of the system.

Nevertheless, such behavior of topological state may not be fulfilled, such in a transverse magnetic field, graphene exists in the topological insulator state with the Chern number  and Hall conductance equal to zero \cite{a3,K3}. The topology of this state is determined by the zero energy Majorana states localized at the boundaries \cite{a3,K3}. We show that this behavior of the system is not unique; the gapless state of the two-dimensional fermion liquid with the  spin-orbit coupling and on-site Coulomb repulsion  can be topological with the zero energy Majorana states.

In topological insulator state the interaction between fermions kills the chiral edge modes \cite{K3,K1} and  a fermion liquid transforms to non-topological insulator state. At the Mott phase transition the repulsion between fermions leads to a gapped spectrum of quasi-particle excitations, the gap opens and insulator state is realized. At the same time the spin-orbit coupling opens a window for topological state formation. The Mott phase transition  occurs as a result of the competition of these processes. We shall show, that in two dimension structures the spin-orbit coupling leads to the Mott transition between topological gapless state and non-topological gapped state. The zero energy Majorana states determine the gapless state  with the Chern number and Hall conductance equal to zero. Thus we study a new type of metal-insulator phase transition.

\section{Model and method}

Consider the Hubbard model on a square lattice, the Hamiltonian of which includes the Rashba spin-orbit coupling ${\cal H}={\cal H}_{Hub}+{\cal H}_{SO}$.
The first term in  ${\cal H}_{Hub}$ is a nearest neighbor hopping term, the second takes into account the on-site repulsion of spinless fermions
\begin{eqnarray}
{\cal H}_{Hub}= -
\sum_{j=1}^{N-1}\sum_{\alpha=1,2}t_\alpha(c^\dagger_{j, \alpha} c_{j+1,\alpha}+ c^\dagger_{j+1, \alpha} c_{j, \alpha})
+
 U\sum_{j=1}^{N}n_{j,1}n_{j,2}.
 \label{eq:H0}
 \end{eqnarray}
 The term ${\cal H}_{SO}$ is defined for the two-orbital square lattice model \cite{1}
 \begin{eqnarray}
&&{\cal H}_{SO}= v \sum_{m,n}[c^\dagger_{n+1,m;1}c_{n,m; 2}- c^\dagger_{n+1,m; 2}c_{n,m; 1}+\nonumber\\&&
 i(c^\dagger_{n,m+1;1}c_{n,m;2}+ c^\dagger_{n,m+1;2}c_{n,m;1})+  H.c.],
\label{eq:H1}
\end{eqnarray}
where $c^\dagger_{j,\alpha},c_{j,\alpha}$ are the fermion operators determined on a lattice site $j=\{n,m\}$ with orbital index $\alpha =1,2$,
$U$ is the  value of the on-site Hubbard interaction determined by the density operator $n_{j,\alpha}=c^\dagger_{j,\alpha}c_{j,\alpha}$,
the band widths of fermions is determined by the hopping integrals $t_\alpha$, where $t_1 = 1$ and $t_2 =t$
($t$ is changed in the interval $0< t \leq 1 $),  $v$ defines the strength of the Rashba  spin-orbit coupling, N is the total number of atoms. As a rule the Rashba spin-orbit coupling is determined by the orbital indices $s$ and $p$, we used the two-digit index $\alpha$.

In the Hubbard model at half filling, a gap opens at the Mott-Hubbard phase transition point. It was shown in \cite{4}, that this phase transition is similar to the Peierls transition. The effective $\lambda$-field is determined by an unknown phase, its value corresponds to the minimum of the action.

Define the operators $\chi_{\textbf{j}}^\dagger= c^\dagger_{\textbf{j}, 2}c_{\textbf{j},1}$ and $\chi_{\textbf{j}}$,  using the Hubbard-Stratonovich transformation for the on-site Coulomb repulsion, we obtain the $\lambda$-field, conjugate with these operators, $\lambda_\textbf{j} \chi^\dagger_\textbf{j}+\lambda^*_\textbf{j}\chi_\textbf{j}$ \cite{4}. The solution for $\lambda_\textbf{j}=\exp(i \textbf{q} \textbf{j}) \lambda $ is determined by an unknown vector $\textbf{q}$ and an amplitude $\lambda$, on which the energies of the quasiparticle excitations depend. The effective Hamiltonian that determines low energy excitations is defined as follows
\begin{equation}
{\cal H}_{eff}(\textbf{k,q}) = \left(
\begin{array}{cccc}
-\varepsilon_1 (\textbf{k}) & \lambda & w(\textbf{k}) & 0\\
 \lambda^*& -\varepsilon_2 (\textbf{k+q}) & 0 & w^*(\textbf{k+q})\\
w^*(\textbf{k}) &0 & -\varepsilon_2 (\textbf{k}) &\lambda^* \\
0 &  w(\textbf{k+q})& \lambda & -\varepsilon_1 (\textbf{k+q} )
\end{array}
\right),
\end{equation}
where $\varepsilon_1({\textbf{k}})=-2\sum_{i=x,y} \cos {k}_i$,  $\varepsilon_2({\textbf{k}})=-2 t \sum_{i=x,y} \cos {k}_i$, $\textbf{k}=(k_x,k_y)$,
$w(\textbf{k})=2 v (-\sin k_x+i\sin k_y)$, $\textbf{k}$ and $\textbf{q}$ are the momenta of fermions.

The $\lambda$-field connects the states of fermions with different orbital indexes and the wave vectors $\textbf{k}$ and $\textbf{k+q}$, the effective Hamiltonian (3) takes into account this connection.
The effective action $S$ has the following form
\begin{eqnarray}
\frac{S}{\beta}=\frac{{1}}{{2N}}\sum_{\textbf{k}}\Psi^\dagger (\textbf{k,q})[\partial_\tau + {\cal H}_{eff}(\textbf{k,q})]\Psi (\textbf{k,q})
+\frac{|\lambda|^2}{{U}},
 \label{A2}
\end{eqnarray}
where $\Psi (\textbf{k,q})$ is the four-component fermion field, two allows for a doubled cell.

Due to symmetry of the spectrum, at half filling the chemical potential is equal to zero for arbitrary $\lambda$, $v$ and $\textbf{q}$. According to numerical calculations \cite{4,6} vector $\textbf{q}=(\pi,\pi)$ corresponds to the minimum energy of system or action $S$, its doubles the cell of the lattice, here
\begin{eqnarray}
\frac{S}{\beta}=-\frac{{T}}{{2N}}\sum_{\textbf{k}}\sum_n \sum_{\gamma=1}^{4} \ln [-i \omega_n+E_\gamma(\textbf{k},\textbf{q})]
+\frac{|\lambda|^2}{{U}},
 \label{A2}
\end{eqnarray}
where $\omega_n =T(2n+1)\pi$ are the Matsubara frequencies, four quasiparticle excitations $E_\gamma(\textbf{k},\textbf{q})$ ($\gamma =1,...,4$) determine the fermion states in the  $\lambda$-field. In the saddle point approximation $\lambda$ is the solution of the following equation $\partial S/\partial \lambda =0$.

At $\textbf{q}=\overrightarrow{\pi}$ the spectrum of quasi particle excitations is symmetric with respect to zero energy, it has the Majorana type. Four branches of the spectrum $\pm E_\gamma (\textbf{k)} $ ($\gamma =1,2$) are determined by the following expression
\begin{equation}
E_\gamma(\textbf{k}) = \sqrt{\alpha(\textbf{k})+(-1)^\gamma \beta(\textbf{k})},
\label{eq:H5}
\end{equation}
where $\alpha (\textbf{k})= \frac{1}{2}(\varepsilon_1^2(\textbf{k})+\varepsilon_2^2(\textbf{k}))+|\lambda|^2+|w(\textbf{k})|^2$,
$\beta^2 (\textbf{k})=\frac{1}{4}[\varepsilon_1^2(\textbf{k})-\varepsilon_2^2(\textbf{k})]^2 + [\varepsilon_1^2(\textbf{k})+\varepsilon_2^2(\textbf{k})] [|\lambda|^2 +|w(\textbf{k})|^2]
-2\varepsilon_1(\textbf{k})\varepsilon_2(\textbf{k}) [|\lambda|^2 -|w(\textbf{k})|^2] +|\lambda|^2[w(\textbf{k})+w^*(\textbf{k})]^2$.

In the ground state the equation for $\lambda$ follows from action (5) and  spectrum (6)
\begin{equation}
\frac{2 |\lambda|}{U}=
\frac{1}{2}\sum_{\gamma=1,2}\int d \textbf{k}  \frac {\partial E_\gamma(\textbf{k})}{\partial |\lambda|}.
\label{eq:H6}
\end{equation}

Below we consider separately the cases of the Hubbard model and the model with bands arbitrary along the band width.

\section{The Hubbard model with the Rashba spin-orbit coupling}

The $\lambda$-field breaks spontaneous symmetry of the Hamiltonian (1), the number of fermions in each band is not conserved. This allows one to take into account the scattering between fermions of different bands and the Hubbard interaction in the Hamiltonian (3). The  $\lambda$-field (or $U$) opens  a gap in the insulator state. The  Rashba spin-orbit coupling (2) breaks the time reversal symmetry of the Hamiltonian (1), forming the topological state in a gapless phase. The Rashba spin-orbit coupling  removes spectrum degeneracy, the fermion spectrum splits into four  branches of quasi particle excitations. The gap between high energy branches is equal to $2\lambda$, as it takes place in the Hubbard model in insulator state \cite{4}. The gap $\Delta$ between the low energy branches depends on the values of  $\lambda$ and $v$, it is equal to zero at $|v|\geq v_c$, here $v_c = \frac{\sqrt{\lambda^2 + \lambda\sqrt{ \lambda^2 + 16}}}{2 \sqrt{2}}$. Using the connection between $\lambda$ and $U$ (7), we calculate the ground state phase diagram in the coordinates ($U,v$), see in Fig 1a).
The value of low energy gap tends to $2 \lambda$  at $\lambda\gg v$. At $|v|<v_c$ an insulator state is realized,  at $|v|>v_c$ exists a gapless state (see the behavior of $\Delta$ and $\lambda$ as function of $U$ and $v$ in Fig 1b) and Fig 1c)). We will consider gapless state in detail, because it is special, as shown below.
For illustration let us consider the phase transition at $v=0.2$ and $0<\lambda\leq \lambda_c =0.0784$, for $\lambda_c$ the gap in the spectrum collapses and fermion liquid transforms into gapless state. At $\lambda>\lambda_c$ the gap opens and the Mott-Hubbard phase transition in an insulator state is realized. At $\lambda <\lambda_c$ a particle-hole symmetry protects the gapless phase. A similar behavior of an electron liquid takes place on a hexagonal lattice in a transverse magnetic field at half filling, when the Hamiltonian is a particle-hole symmetric. With another filling, the states of the topological insulator are realized \cite{a3,K3}.

Nontrivial peculiarities in the behavior of the fermion liquid is determined by the low energy branches of the spectrum in a gapless state, therefore we consider the transformation only low energy spectrum at the Mott-Hubbard transition. For doubled cell the interval for wave vectors is limited by $\pi$ ($0\leq k_x, k_y \leq \pi$). At the phase transition point $\lambda_c$, the low-energy branch touches  zero energy in two points in the k-planes $k_x=0$, $k_y=0.4554$ and $k_x=\pi$ , $k_y=\frac{1}{2}-0.4554$  forming   a gapless state (see in Figs 2). Due to the Rashba spin-orbit coupling, in the gapless state a low-energy spectrum is split twice near  zero energy (see in Figs 3   at $\lambda=0.05<\lambda_c$, for illustration). There are two points, that touch zero energy in each $ (0,\pi)$ k-plane: $k_x=0$ $k_y=0.4418$, $k_y=0.4788$ and
$k_x=\pi$ $k_y=0.0212$, $k_y=0.0582$ (see in Fig 3c)). As a rule, the point of the phase transition, in which a gap collapses, separates two different phase. In our case the points in k-planes, in which the gaps are equal to zero, separate topological (between these points) and non topological states in the $k_y$-space regions. Numerical calculations show, that topological state is characterized by the zero energy Majorana states, localized at the boundaries (see in Fig 3d)) \cite{a3,K3}. We consider the 2D fermion system in the stripe geometry with open boundary conditions for the boundaries along the y-direction.
The real part of the wave function of the zero energy Majorana states as function of lattice site along the x-axis is shown in Fig 3d), the wave function is calculated for $\frac{k_y}{2\pi}=0.46$ (for all other wave vectors from these $k_y$-space regions, the behavior of the wave function is the same).
Numerical calculations show, that topological semimetal state exists for an arbitrary $v>v_c$, the k-space regions, in which the zero energy Majorana states are realized, increase with increasing $v$.
Thus, the 2D topological state with the zero energy Majorana states is realized for given regions of wave vectors. Note that, the Chern number and the Hall conductance are zero, this is the state of the 2D topological semimetal (the gap equal to zero) with zero Chern number. For the first time such topological state was obtained in the framework of the Hofstadter model on a honeycomb lattice \cite{a3,K3}.

\begin{figure}[tp]
     \centering{\leavevmode}
\begin{minipage}[h]{.315\linewidth}
\center{
\includegraphics[width=\linewidth]{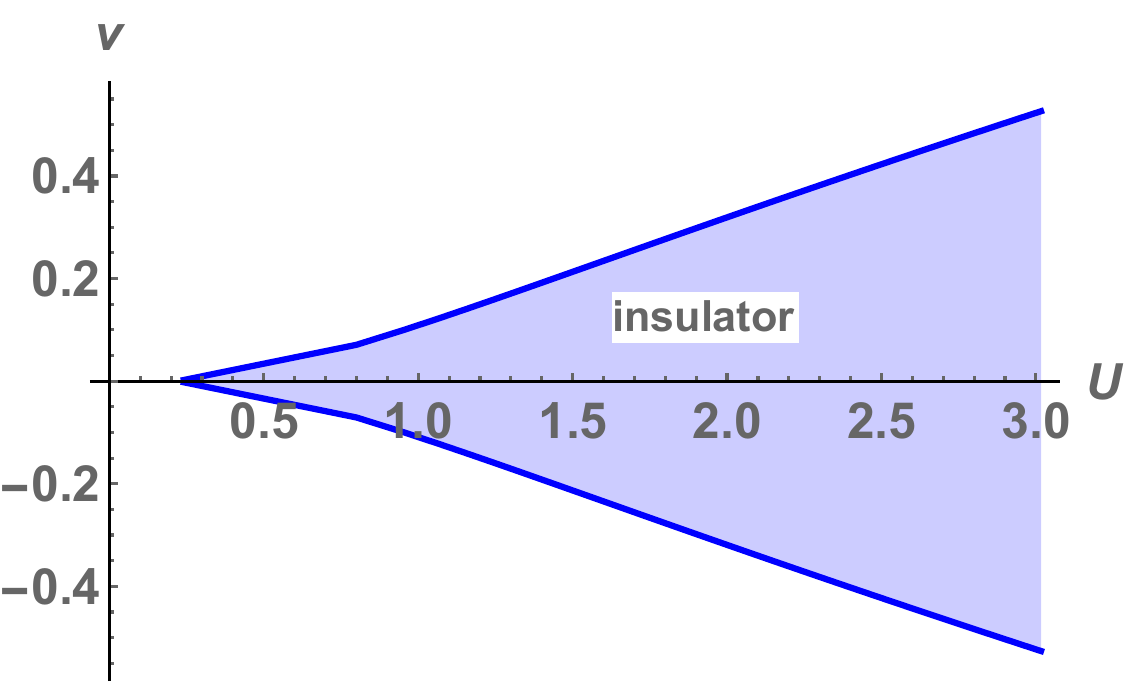} a)\\
                 }
   \end{minipage}
   \centering{\leavevmode}
\begin{minipage}[h]{.315\linewidth}
\center{
\includegraphics[width=\linewidth]{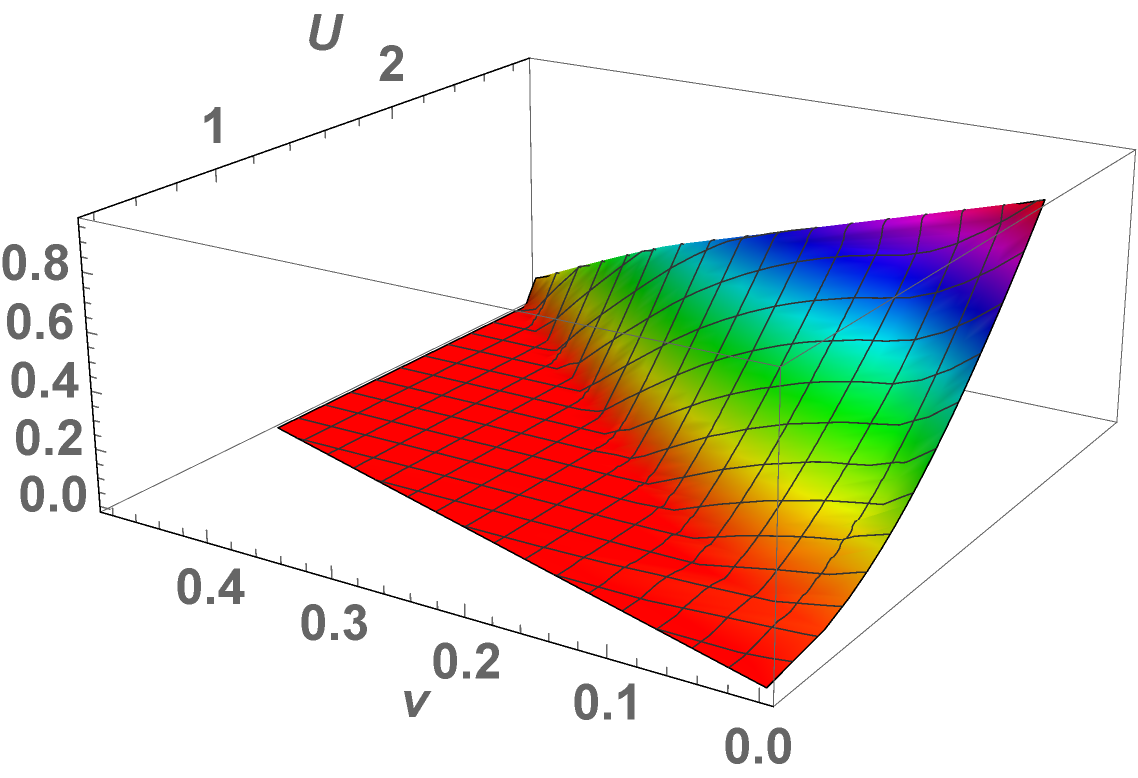} b)\\
                 }
   \end{minipage}
    \centering{\leavevmode}
\begin{minipage}[h]{.315\linewidth}
\center{
\includegraphics[width=\linewidth]{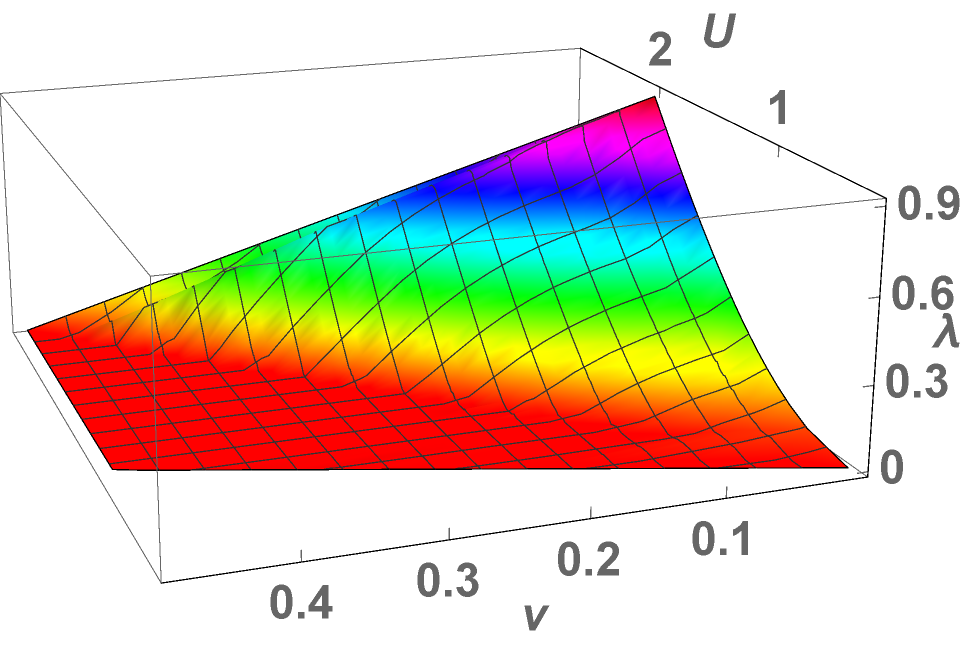} c)\\
                 }
   \end{minipage}
\caption{(Color online)
The ground state phase diagram of the Hubbard model ($t=1$) in the coordinates $(U,v)$ a).
 The gap of low energy quasiparticle excitations b) and $\lambda$ c) as a function of $U$ and $v$.
  }
\label{fig:1}
\end{figure}
\begin{figure}[tp]
     \centering{\leavevmode}
\begin{minipage}[h]{.37\linewidth}
\center{
\includegraphics[width=\linewidth]{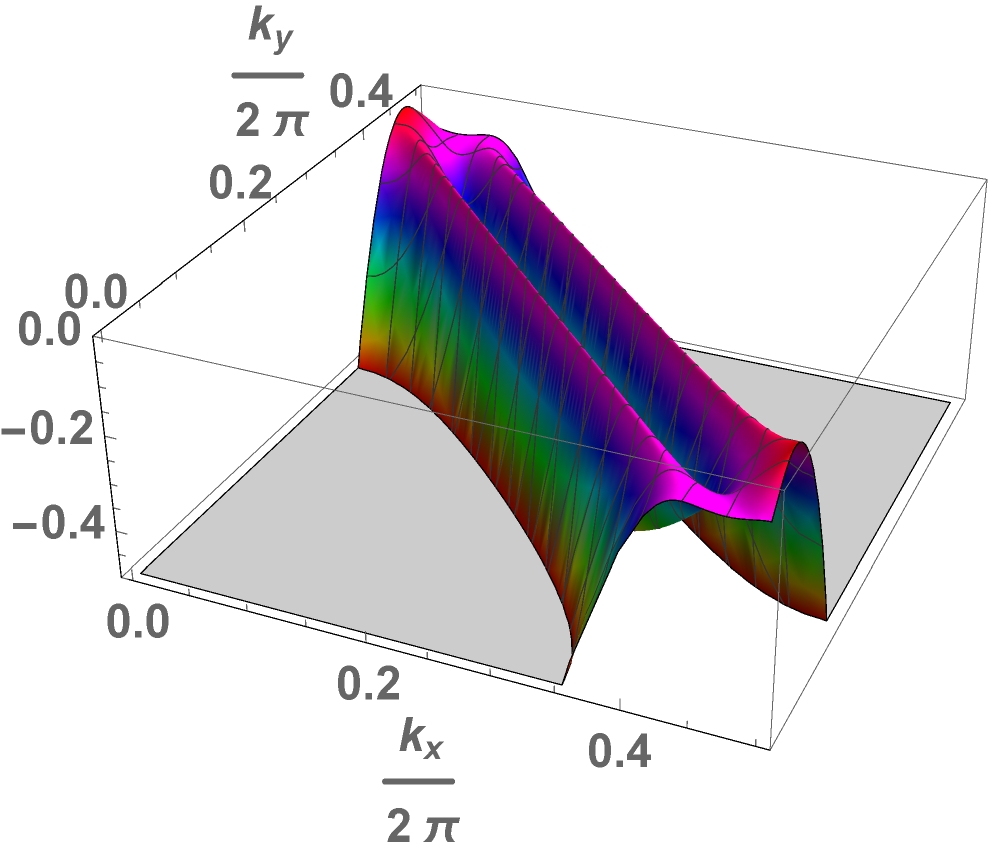} a)\\
                 }
   \end{minipage}
     \centering{\leavevmode}
     \begin{minipage}[h]{.32\linewidth}
\center{
\includegraphics[width=\linewidth]{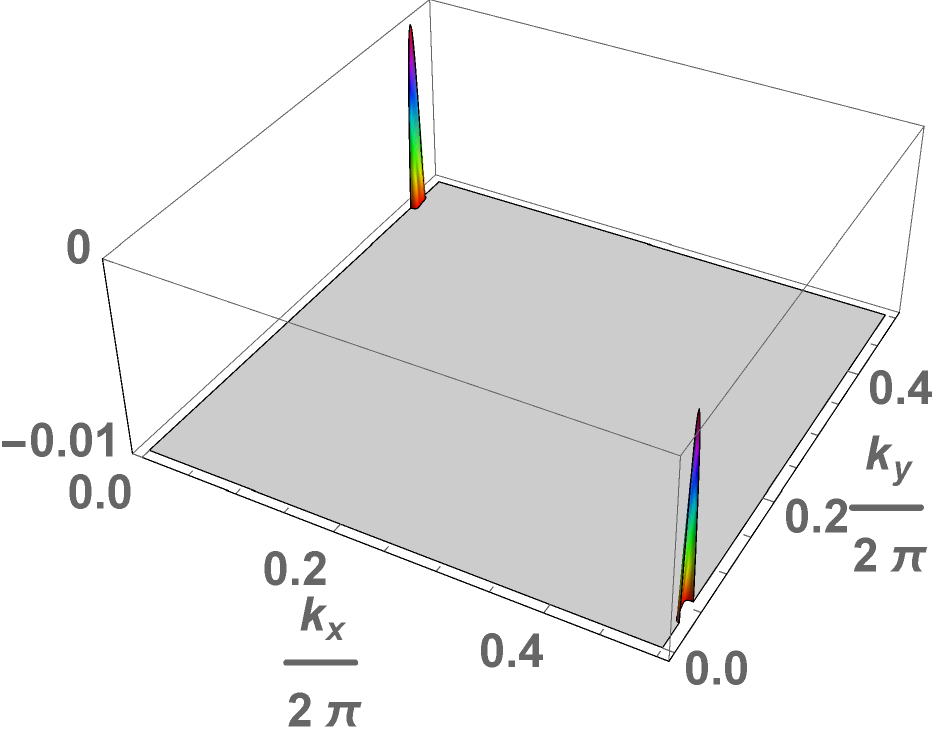} b)\\
                 }
   \end{minipage}
    \centering{\leavevmode}
     \begin{minipage}[h]{.25\linewidth}
\center{
\includegraphics[width=\linewidth]{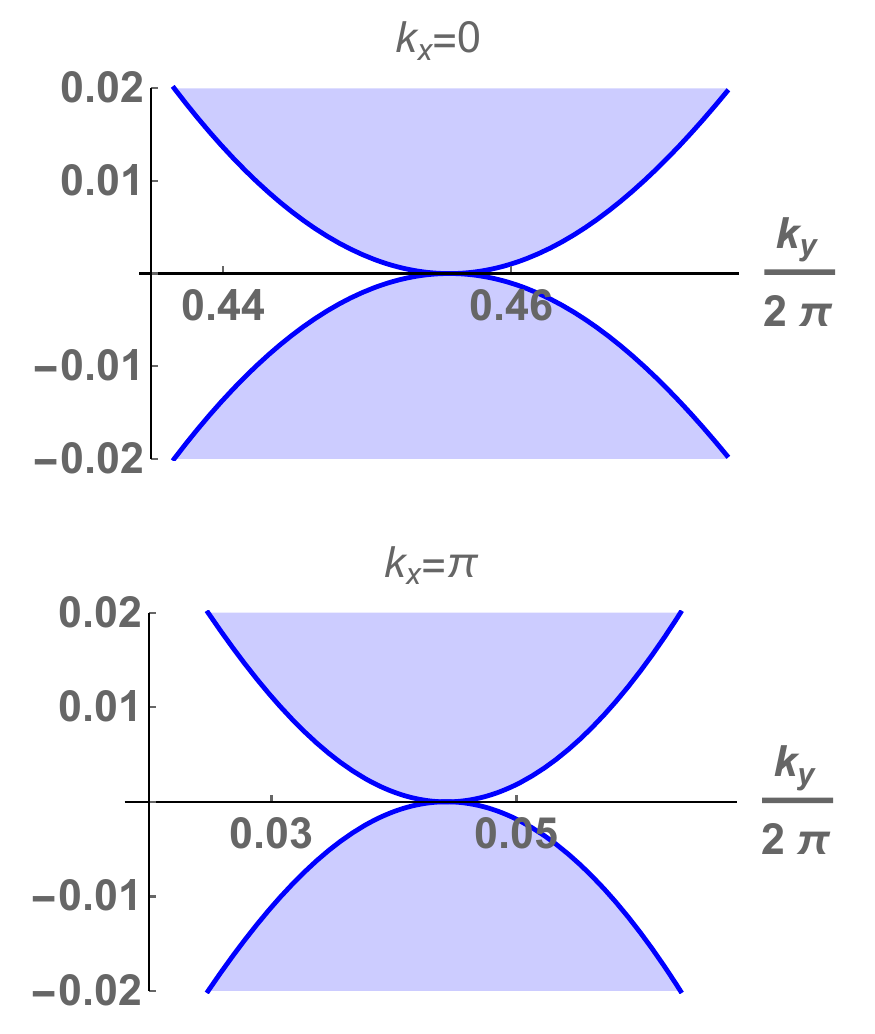} c)\\
                 }
   \end{minipage}
\caption{(Color online)
A low energy branch of the spectrum at the point of the Mott-Hubbard transition calculated at $\lambda_c=0.0784$, $v=0.2$; a low energy part of the spectrum a), near the zero energy b), c) cutting at $k_x=0$ and $k_x=\pi$.
  }
\label{fig:2}
\end{figure}
\begin{figure}[tp]
     \centering{\leavevmode}
\begin{minipage}[h]{.37\linewidth}
\center{
\includegraphics[width=\linewidth]{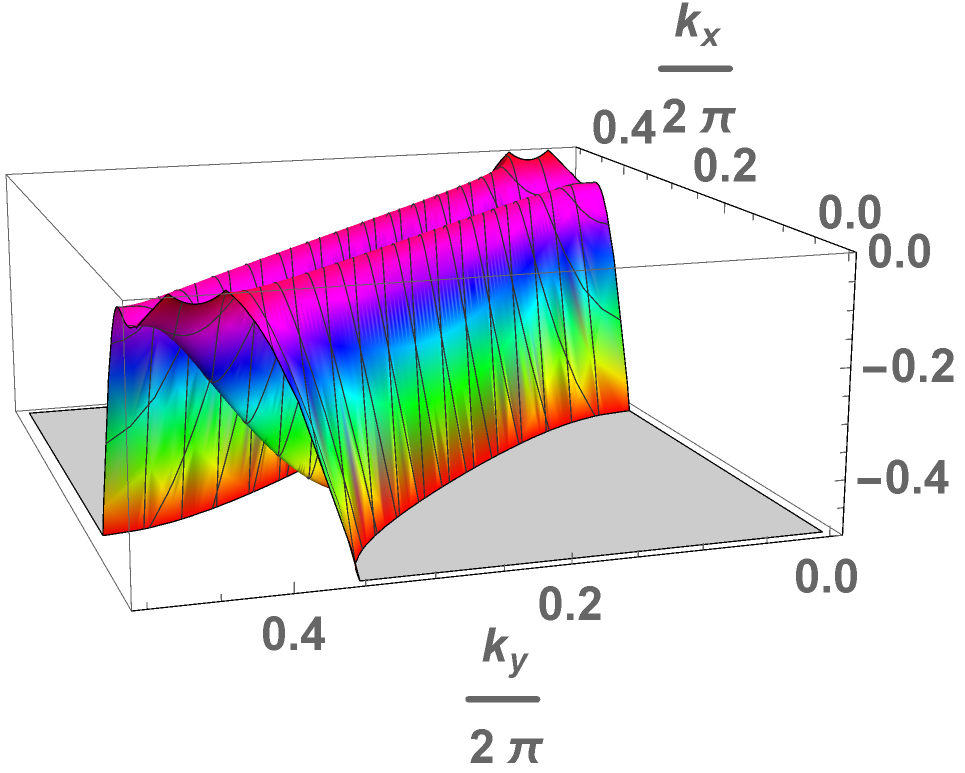} a)\\
                 }
   \end{minipage}
     \centering{\leavevmode}
     \begin{minipage}[h]{.32\linewidth}
\center{
\includegraphics[width=\linewidth]{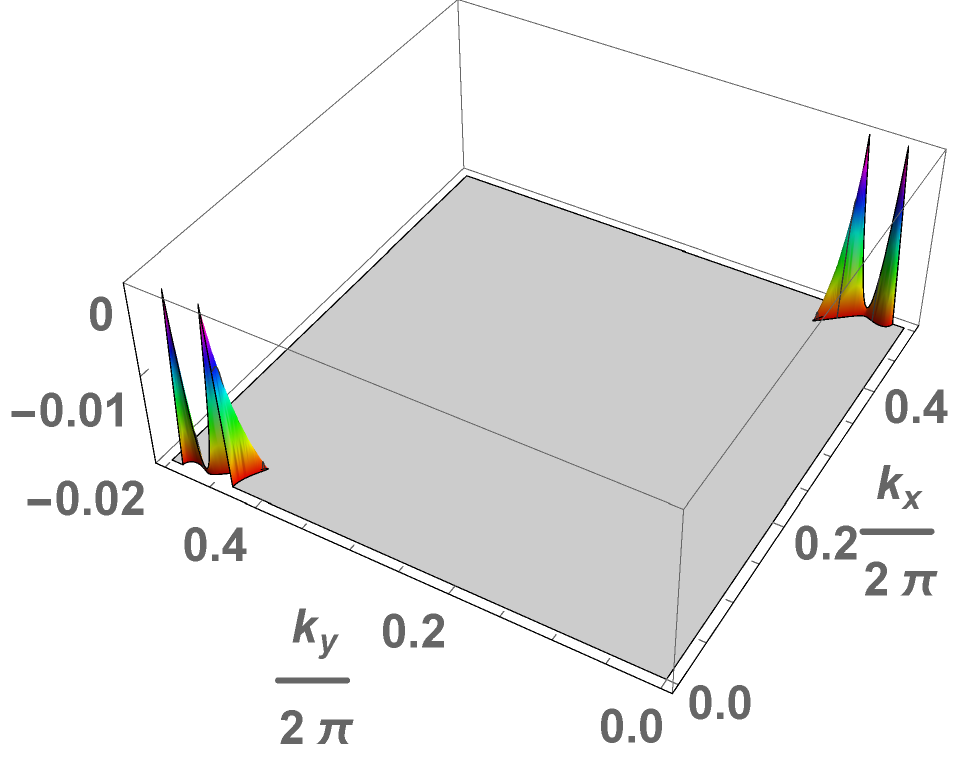} b)\\
                 }
   \end{minipage}
    \centering{\leavevmode}
     \begin{minipage}[h]{.27\linewidth}
\center{
\includegraphics[width=\linewidth]{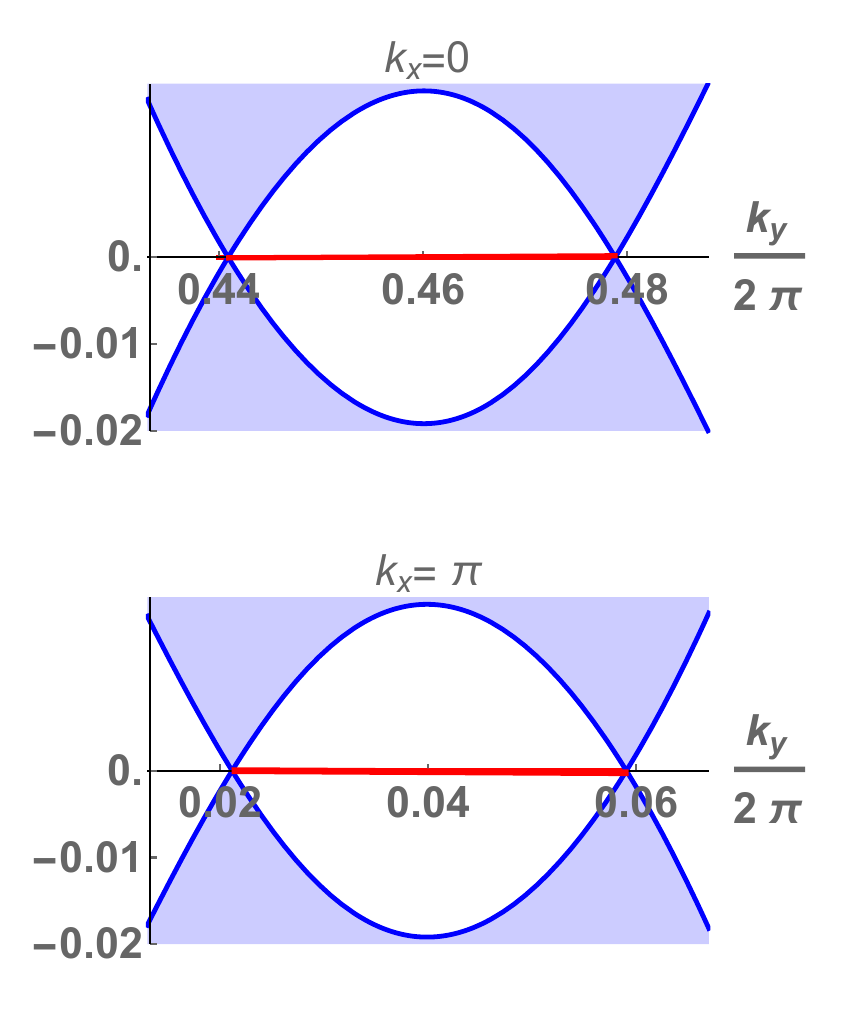} c)\\
                 }
   \end{minipage}
   \centering{\leavevmode}
     \begin{minipage}[h]{.47\linewidth}
\center{
\includegraphics[width=\linewidth]{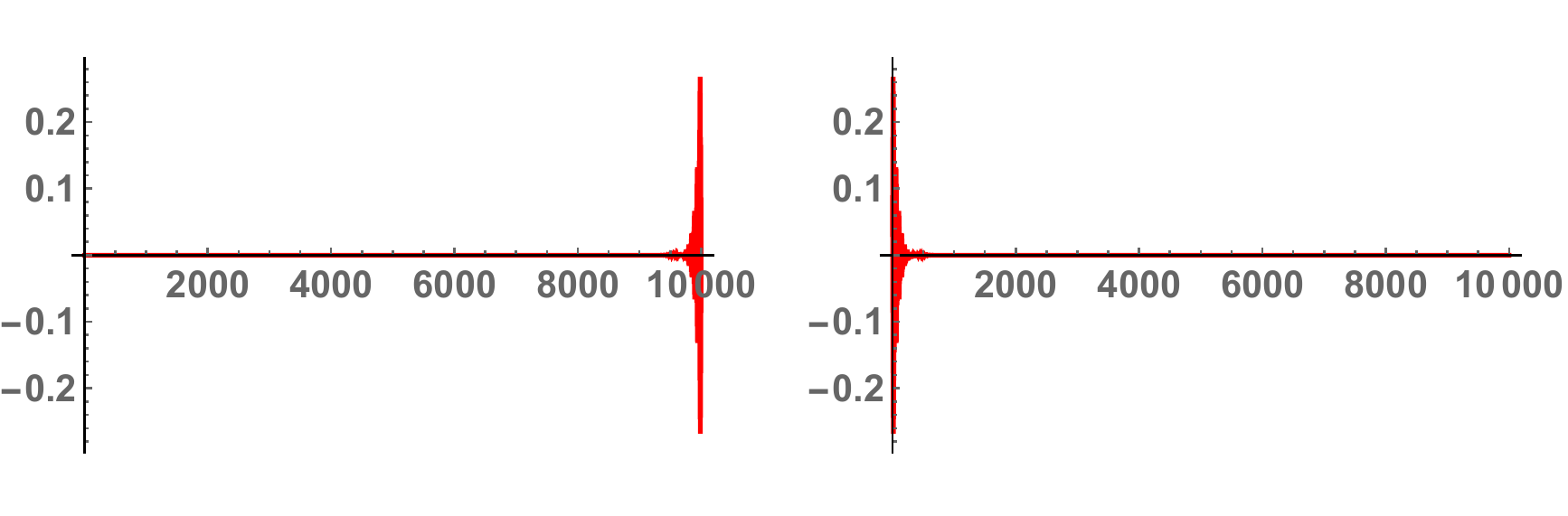} d)\\
                 }
   \end{minipage}
\caption{(Color online)
A low energy branch of the spectrum in gapless phase calculated at $\lambda=0.05$, $v=0.2$; a low energy part of the spectrum a), near zero energy b), cutting at $k_x =0$ and $k_x =\pi$ c) (the regions of existence of band states and zero energy Majorana states are highlighted in blue and red,
respectively),  the wave function of the zero energy Majorana states as a function of coordinate $1\leq x \leq 10^4$ (x=1 and $x=10^4$ are the boundaries), calculated at $\frac{k_y}{2\pi}=0.46 $ d).
  }
\label{fig:3}
\end{figure}

\section{The two band model with the Rashba spin-orbit coupling}

According to the ground state phase diagram, presented in Fig 1a), the  region of existence of the insulator state increasing with $U$ and decreasing with $v$. As we shown in previous section, at $t=1$ the gapless state is topological. At $0<t\leq 1$ the fermion liquid is realized in the same phase state for arbitrary $t$ at given $U$ and $v$. A critical value of $v$, at which the gap opens, is equal to $v_c = \frac{\sqrt{\lambda^2 + \lambda\sqrt{ \lambda^2 + 16 t}}}{2 \sqrt{2}}$. We do not discuss the point $t=0$, when the Hamiltonian (1) describes the Falicov-Kimball model. These are two points of view about the behavior of fermion liquid in the  Falicov-Kimball model  at half-filling:  popular, that this is an exciton insulator state \cite{D} and unpopular, that this is a point of the phase transition exhibiting in the model (1) at $t=0$ \cite{4}. At $ t = 0 $, the phase of the $ \lambda $ -field is not defined; it changes jumpwise from $\pi$ to 0. Within the framework of the considered approach, we do not know the behavior of the fermion liquid (1) at $ t = 0 $. In this context, it makes no sense to consider the Mott phase transition in (1),(2) at $t=0$, when taking into account the  Rashba spin-orbit coupling.
\begin{figure}[tp]
     \centering{\leavevmode}
\begin{minipage}[h]{.4\linewidth}
\center{
\includegraphics[width=\linewidth]{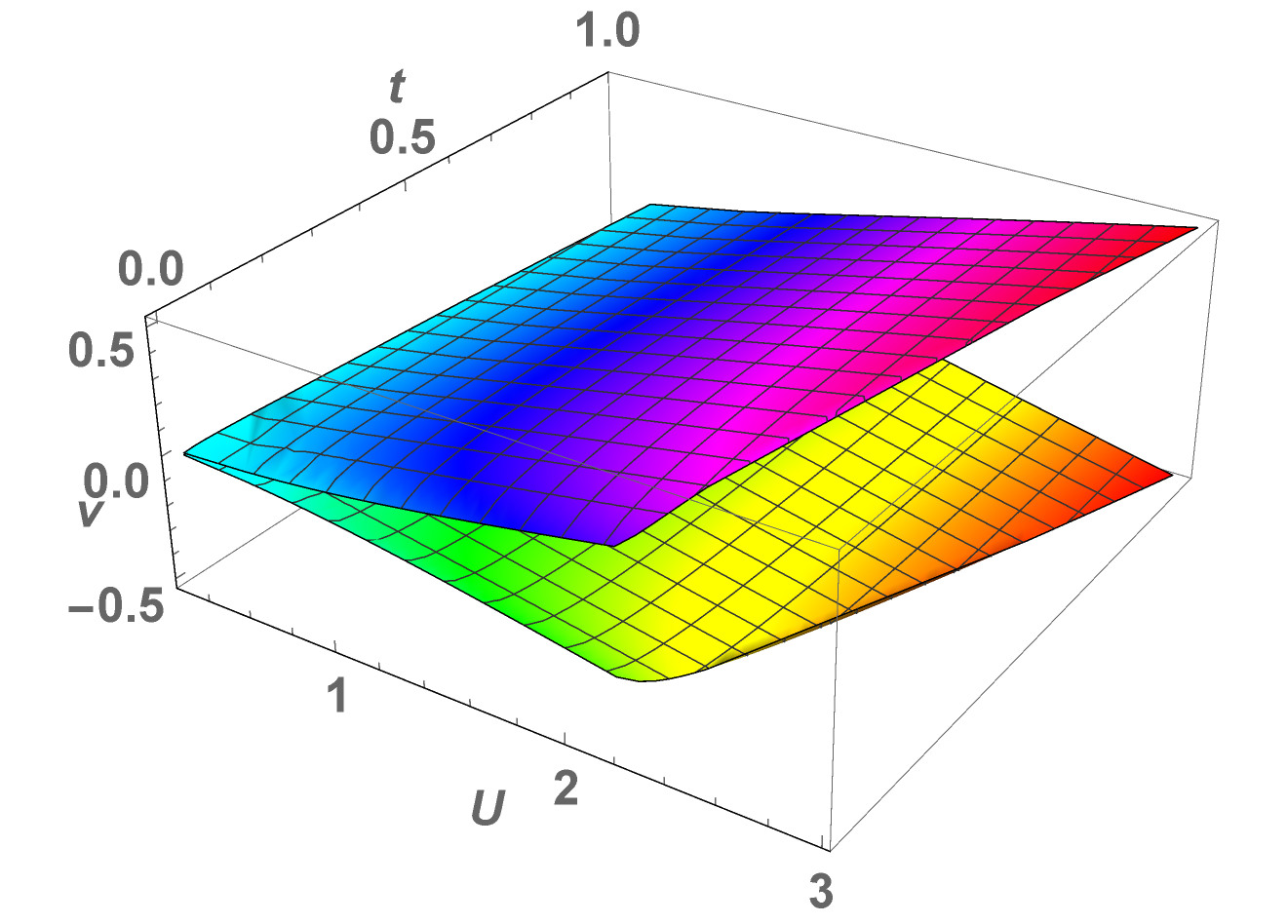} a)\\
                 }
   \end{minipage}
     \centering{\leavevmode}
\begin{minipage}[h]{.4\linewidth}
\center{
\includegraphics[width=\linewidth]{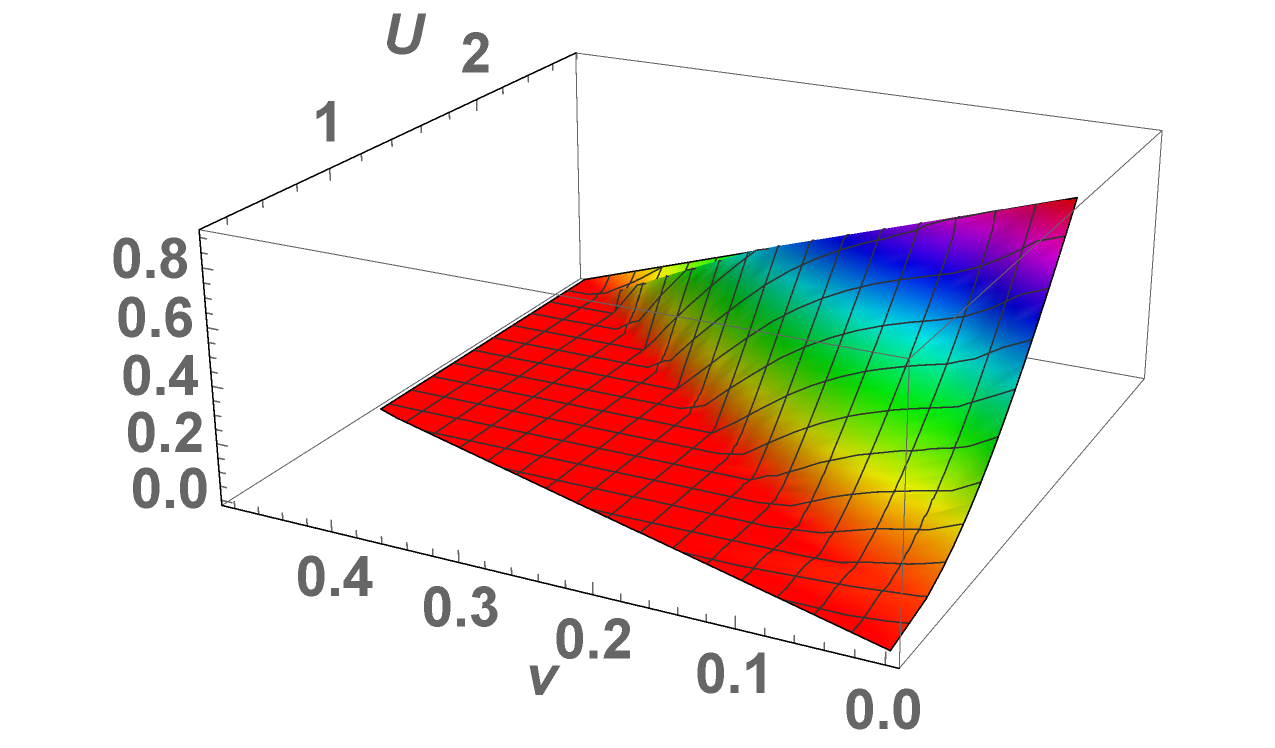} b)\\
                 }
   \end{minipage}
\caption{(Color online)
The ground state phase diagram of the model with different band widths ($0<t<1$) in the coordinates $(U, t, v)$ a), the insulator phase is limited by plates. The gap of low energy quasiparticle excitations as a function of $U$ and $v$, calculated at $t=\frac{1}{2}$ b).
  }
\label{fig:4}
\end{figure}
\begin{figure}[tp]
     \centering{\leavevmode}
\begin{minipage}[h]{.38\linewidth}
\center{
\includegraphics[width=\linewidth]{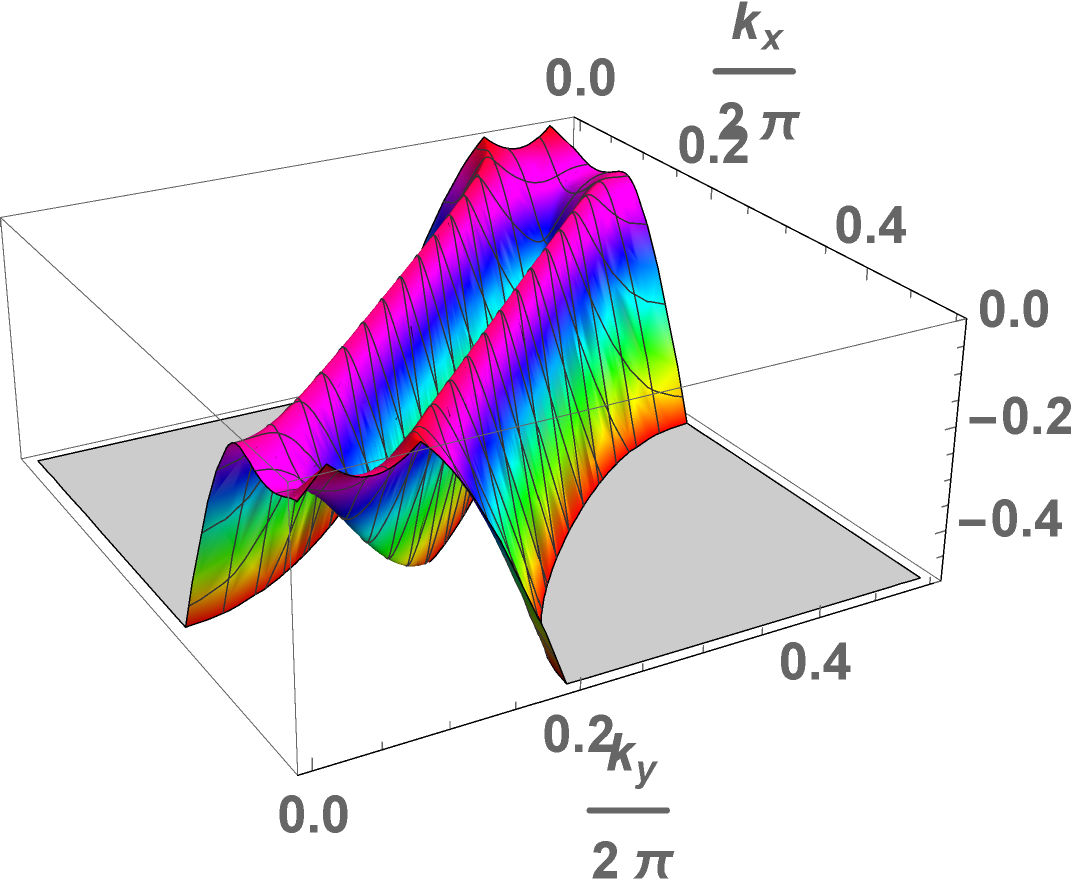} a)\\
                 }
   \end{minipage}
     \centering{\leavevmode}
\begin{minipage}[h]{.33\linewidth}
\center{
\includegraphics[width=\linewidth]{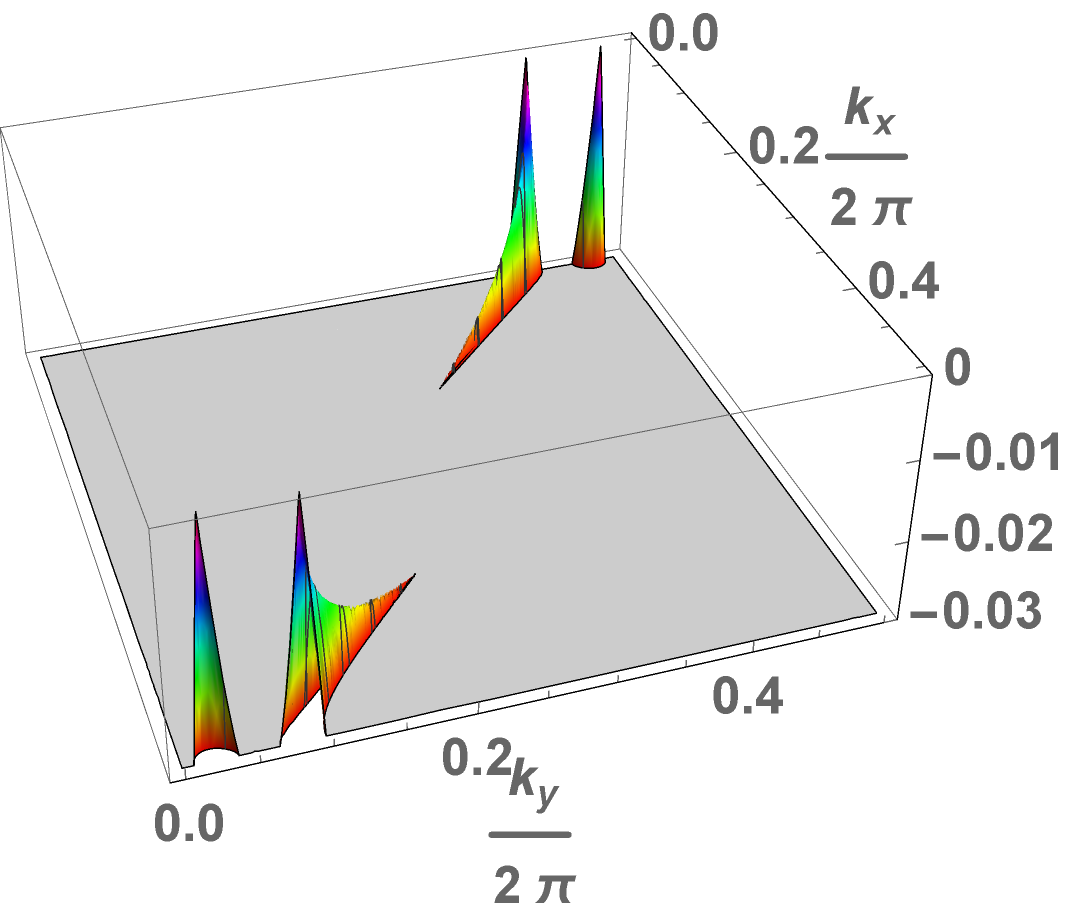} b)\\
                 }
   \end{minipage}
    \centering{\leavevmode}
\begin{minipage}[h]{.25\linewidth}
\center{
\includegraphics[width=\linewidth]{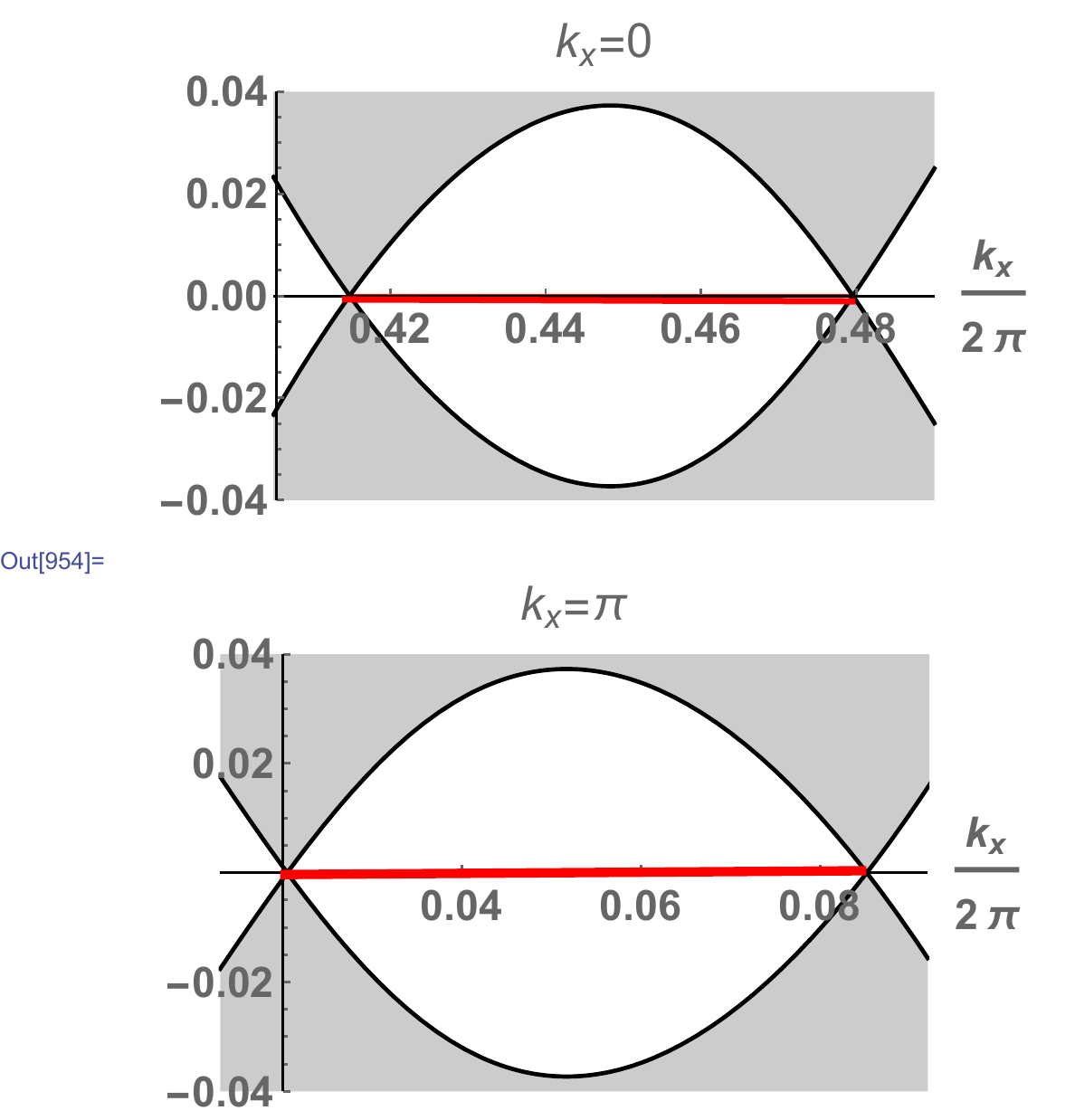} c)\\
                 }
   \end{minipage}
\caption{(Color online)
A low energy branch of the spectrum in gapless phase calculated at $\lambda=0.05$, $v=0.2$,  $t=\frac{1}{2}$; a low energy part of the spectrum a), near zero energy b), cutting at $k_x =0$ and $k_x =\pi$ c) (the regions of existence of band states and zero energy Majorana states are highlighted in blue and red
respectively). }
\label{fig:5}
\end{figure}

The value of $v_c$ increases with increasing $t$, the region of existence of gapless states decreases (see in Fig 4a)). At critical value of $v_c$ the gap opens (see in  Fig 4b)), $k$-space regions in which, the Majorana zero energy states are realized, increases with decreasing $t$. We consider an intermadiate value of $t=\frac{1}{2}$ at the same $\lambda=0.05$ and $v=0.2$. The details of numerical calculations of a low energy spectrum are shown in Figs 5. The value of the splitting of the spectrum near zero energy increases with decreasing $t$. In Fig 5c) the k-space regions of existence of zero energy Majorana state are shown, where  at $k_x=0$ $k_y=0.4149$, $k_y=0.4795$ and at $k_x=\pi$ $k_y=0.0205$, $k_y=  0.08519$. These state are localized at the boundaries, the wave function has the form  as in Fig 3d).

\section{Conclusions}
The two-dimensional model of spinless fermions with different band widths  is considered at half filling within the mean-field approach.
The Mott-Hubbard phase transition in the Hubbard model was also studied as a special case. The ground state phase diagram  of the model is calculated.
It is characterized by two phase states: an insulator and a topological semimetal. Note, that the point of the Mott phase transition is realized at  not strong spin-orbit coupling, such  in the Hubbard model when  $0.05<v<0.2$,  $\frac{v}{U}<0.28$.
It is shown that the nontrivial topology of the gapless phase is due to the zero energy Majorana states, localized at the boundaries of the sample. These states are dispersionless chiral modes in  certain ranges of the one-dimensional wave vector. It was shown in Refs \cite{a3,K3} that such  topological state is realized in graphene in an external magnetic field applied perpendicularly. Despite the fact that we are talking about the two-dimensional model, the Chern number and  the Hall conductance are equal to zero. When taking into account the Rashba spin-orbit coupling in two dimensional systems, the Mott phase transition must be considered as a phase transition between the topological semi-metal and non-topological insulator states.


\end{document}